# Graphene Physics in Graphite

By Yakov Kopelevich[*] and Pablo Esquinazi[**]

*Single layers of carbon dubbed "graphenes", from which graphite is built, have attracted broad interest in the scientific community because of recent exciting experimental results. Graphene is interesting from a fundamental research perspective, as well as for potential technological applications. Here, we provide a brief overview of recent developments in this field, focusing especially on the electronic properties of graphite. Experimental evidence indicates that high-quality graphite is a multi-layer system with nearly decoupled 2D graphene planes. Based on experimental observations, we anticipate that thin graphite samples and not single layers will be the most promising candidates for graphene-based electronics.*

I. **Introduction.**

Graphite is the oldest known allotrope of crystalline carbon. In the last 20 years, discoveries of quasi-0D buckyballs (fullerenes)[1] and quasi-1D graphene sheets wrapped into cylinders (carbon nanotubes (CNTs))[2] have resulted in renewed interest in the properties of graphite. However, recent evidence of phenomena such as i) a magnetically driven metal-insulator transition (MIT),[3] ii) the quantum Hall effect(QHE),[4] and iii) the existence of 2D Dirac fermions (DFs)[5-7] in graphite clearly indicate that much of the physics of this material has been missed in the past. Although the fabrication of graphite-based mesoscopic devices with dimensions of a few tens of nanometers has been reported previously in 2001,[8] the discovery of novel physical properties has triggered research into few-layer graphite (FLG) samples and individual graphitic sheets.[9-11]

The technique used for the preparation of FLG samples, involving the mechanical extraction of FLG sheets from bulk highly oriented pyrolytic graphite (HOPG) or Kish-graphite samples, is at first glance surprisingly simple, and is based on the extremely weak interlayer interaction between graphene planes. Using this method, 2D graphite quantum dots[12] and electromechanical resonators have been fabricated from multilayer graphite and graphene.[13] Additionally, proximity-induced superconductivity has been observed in both FLG and graphene.[14] Remarkably, the QHE has been observed at room temperature in these samples.[15] All this work suggests that graphite/graphene may be a promising material for micro- and nanoelectronics.

Clearly, it is important to develop better understanding of the physical properties of graphene for application purposes. In this article, we focus on results demonstrating the 2D behavior of quasi-particles (QPs) in bulk graphite, especially multiple layers of graphene and FLG samples. We also comment on some of the important experimental similarities and differences between these systems.

Graphite consists of several layers of honeycomb lattices of carbon atoms, characterized by two non-equivalent sites, A and B, in the Bernal (ABABAB...) stacking configuration.[16] In the absence of interlayer electron hopping, the Fermi surface (FS) is reduced to two points (K and K′) at the opposite corners of the 2D hexagonal Brillouin zone (BZ) where the valence and conduction bands touch each other, leading to a linear dispersion relation $E(p) = \pm v|p|$ (where $E$ is the energy and $p$ is the QP momentum) for p-electrons that move parallel to but not within the graphene layers. Hence, the carriers can be described as massless (2+1)D DFs,[17] thus providing a link to relativistic models for particles with an effective "light" velocity $v \approx 10^6$ ms$^{-1}$.

One question that remains is that if the interlayer electron coupling in graphite is not zero, exactly how strong is it ? There is still no clear answer to this fundamental question, primarily because of a rather inelegant reason: no measurements have been performed on defect-free graphite samples.

According to the frequently used Slonczewski-Weiss-Mc-Clure (SWMC) model,[16] interlayer coupling dramatically modifies all properties of the electron gas in graphite, leading to a dispersion in the direction perpendicular to the ($p_x$, $p_y$) planes with cigar-like FS pockets elongated along the corner edge HKH of a 3D BZ. The SWMC model uses or estimates an overlap integral $\gamma_1 \approx 390$ meV between the nearest layers. However, this value of $\gamma_1$ is nearly two orders of magnitude larger than the value of ca. 5 meV reported previously in the literature by Haering and Wallace,[18] who have pointed out the 2D character of QPs in graphite. We would like to point out that the relationship between the interlayer binding energy per carbon atom and $\gamma_1$ is model dependent and has still not been adequately resolved.

One of the possible reasons for the uncertainty in experimentally determining the binding energy perpendicular to the graphene layers is the presence of lattice defects in graphite that can act essentially as short-circuits between the planes, thereby inducing a 3D character to the measured FS with a new QP density of states at the Fermi level. Defects also act as effective doping centers thus changing the QP density. Indirect evidence for this general behavior is provided by the spread of values for the out-of-plane/basal-plane resistivity ratio $\rho_c/\rho_b$ , which reaches values greater than $10^4$ at room temperature[4] for oriented graphite samples with mosaicity $\leq 0.3°$, indicating the weak overlap of p-electron wave functions in the $c$-axis direction, but decreases to a value below 100 for samples such as Kish graphite with mosaicity >1°, sometimes erroneously called "single crystals" in the literature. Perhaps the latest reported experimental value for the interlayer cohesive or binding energy per carbon atom in graphite is 52 ± 5 meV, which has been measured for a low-quality, grade ZYB HOPG sample.[19]

## 2. 2D Electrical Behavior of Bulk Graphite: The QHE

According to the SWMC model, coherent transport is expected for the interlayer magnetoresistance $\rho_c(B,T)$ at low temperatures. Indeed, high-resolution angle-dependent measurements of $\rho_c(\theta)$ reveal a maximum (the so-called "coherent resistivity peak") when the magnetic field is applied parallel to the graphene layers for less-ordered graphite samples with mosaicity >1°.[20] In contrast, for well-ordered HOPG samples with mosaicity ≤0.4°, no maximum has been observed. This result suggests incoherent transport along the $c$-axis, and therefore the existence of a 2D FS in "ideal" graphite. Indeed, it is such highly ordered graphite samples that exhibit the QHE, which is characterized by plateau-like steps in the Hall resistance $R_{xy}(B)$, which is in clear contrast to the absence of plateaus in more disordered graphite samples.[4] Figure 1 shows examples of the Hall conductance ($G_{xy} = 1/R_{xy}$) data[4] obtained for two HOPG bulk samples with contacts placed at the sample surface. Here, the reduced Hall conductance $-G_{xy}/G_{0xy}$ is plotted as a function of the filling factor $B_0/B$, where $G_{0xy}$ corresponds to the step between subsequent QHE plateaus and the normalization field $B_0$ = 4.68 T[5,21] corresponds to the Shubnikov-de Haas (SdH) oscillation frequency, which is proportional to the 2D QP density.

A comparison of the data with the expected behavior for massive QPs as well as DFs indicates contributions from both in HOPG-UC, whereas for the more disordered HOPG-3 sample, massive QP contributions to the Hall resistance seem to be predominant. We have previously reported an algorithm to separate the contributions from Hall and longitudinal resistances in the experimental data.[21] We stress that plateau-like features have been observed in the Hall conductance of strongly anisotropic graphite samples, which lack the so-called "coherent resistivity peak" and hence a 3D FS.[20] In other words, these samples exhibit an out-of-plane/basal-plane resistivity ratio $\rho_c/\rho_b > 10^4$ originating from the high degree of crystallite orientation along the hexagonal c-axis, as confirmed by X-ray rocking curve measurements. In contrast, in quasi-3D Kish graphite with $\rho_c/\rho_b \sim 100$, the mosaicity exceeds 1.4°, and no signatures for the Hall plateaus have been detected.[4]

The behavior shown in Figure 1 provides clear experimental evidence for the occurrence of the QHE in graphite. However, by comparing these results with typical Hall resistivity data for 2D electron gas systems, we note that in HOPG the longitudinal resistivity $\rho_{xx} \equiv \rho_b$ neither goes to zero nor shows clear minima accompanying the plateaus at the corresponding filling factors;[4] indeed, this value appears to exceed the Hall resistivity $\rho_{xy}$. Apparently, the inequality $\rho_{xx} > \rho_{xy}$ holds in the QHE regime of HOPG, which can be attributed to the significant amount of structural disorder existing even in the best quality samples. We speculate that this disorder is also the origin of the anomalous linear field dependence and lack of saturation of $\rho_{xx}(B)$ at high fields and low temperatures.[4,22] It is important to point out that similar to HOPG, the QHE in Bechgaard salts,[23] as well as in more conventional 2D electron gas system like GaAs/AlGaAs,[24] has also been observed in the regime where $\rho_{xx} > \rho_{xy}$.

Since two kinds of massive QPs have been detected in graphite,[5,7] specifically massive conventional fermions with Berry's phase zero and "chiral" QPs with Berry's phase $2\pi$, additional analysis of the QHE data is needed (the term "chirality" implies that pseudospin (sublattice index) is associated with the momentum of the QP. The spinning of these "chiral" QPs in an applied magnetic field leads to the pseudospin rotation that introduces a phase shift in the QP wavefunction, i.e., Berry's phase). QPs with Berry's phase $2\pi$ have been theoretically predicted[25] and observed experimentally[26] for bilayer graphene samples, and are massive QPs with a touching twoparabola spectrum. Unlike conventional QPs that follow the energy relation $E(p) = p^2/2m_e^*$, the massive "chiral" QPs obey $E(p) = \pm p^2/2m_e^*$, where $m_e^*$ is the effective electron mass. Their quantized energy levels in a magnetic field are given by two different relations: $E_n = \hbar\omega_c(n + \frac{1}{2})$ and $E_n = \pm \hbar\omega_c[n(n-1)]^{1/2}$.[25] Here, $\omega_c = eB/m_e^*$, n is the Landau level (LL) number (=0, 1, 2,...), and $\hbar$ is the Planck constant divided by $2\pi$. Upon plotting the reduced Hall conductance as a function of the filling factor (Fig. 1), staircase 1 for conventional fermions and staircase 2 for massive chiral QPs are expected. There is an unconventional doubly degeneracy at the lowest Landau levels (n = 0, 1) but for higher n the system recover the behavior for normal carriers.[21] Measurements at the lowest Landau levels are needed to distinguish between these two contributions. Since the contribution of massive electrons to the electronic properties of graphite depends on the defect concentration and the overall sample quality, it may be reasonable to believe that the massive QPs are not necessarily intrinsic but depend on the actual coupling between layers and the overall sample quality. Certainly, a clear answer to this question requires further experimental work.

A recent theoretical model of the 3D QHE in graphite,[27] based on the SWMC model, predicts the occurrence of only one plateau in the Hall conductivity $\sigma_{xy}(B)$ for applied magnetic fields $B > B_{QL}$, where $B_{QL} \sim 8$ T is the field that pulls all carriers into the lowest LL. However, as illustrated by Figure 1, various Hall plateaus are observed for $B < B_{QL}$ in both the experimentally measured HOPG samples, providing further evidence for the 2D behavior of graphite. It is therefore reasonable to conclude that the SWMC model is not very well suited for describing the transport properties of high quality graphite.

To emphasize the similarities between bulk HOPG and FLG, we have plotted in Figure 1 the Hall data reported for FLG at a bias voltage of 80 V;[9] these results agree nicely with the results obtained previously for HOPG.[4] Note that the normalization constant for FLG,[9] $B_0 = 20$ T, is much larger than the corresponding value for high-quality HOPG samples. This is because of the larger defect density, which effectively leads to doping of the graphene lattice. Owing to the method of fabrication, as well as the possible effects of surface doping, the FLG samples exhibit much higher carrier density in general at zero bias voltage as compared to HOPG samples; the FLG samples also show less clear plateaus in the Hall data.

Other experimental evidence corroborating the 2D behavior of HOPG comes from scanning tunneling spectroscopy (STS) and microscopy (STM) experiments.[7,28,29] In particular, Li et al. have shown the presence of massive QPs in HOPG, which coexist with the DF. This result agrees nicely with the results obtained from quantum oscillations and QHE measurements.[4,5,21]

## 3. Dirac Fermions in Graphite

The scientific community has been especially focused on the behavior of DFs owing to their linear dispersion relation and expected relativistic-like behavior. The first unambiguous experimental evidence for DFs in bulk graphite has been obtained from quantum oscillation measurements.[5] Thus, the oscillating longitudinal conductivity (SdH effect) in a quasi-2D system is $\sigma_{xx} \sim -\cos[2\pi(B_0/B - \gamma)]$ with a phase factor $\gamma = \frac{1}{2}$ or $\gamma = 0$ for normal massive and Dirac QP, respectively. The $\gamma = 0$ phase value corresponds to the Berry's phase $\pi$, which has also been detected in SdH and QHE measurements performed on graphene.[10,11] Moreover, there is solid experimental data to evidence the existence of DFs in bulk graphite as well as FLG. Angle-resolved photoemission spectroscopy (ARPES) measurements show a linear $E(p)$ spectrum in the vicinity of the BZ corner H,[6] and STS experiments[7] further corroborate the occurrence of DFs in bulk graphite samples via direct measurements of the LL quantization spectrum $E_n = \pm (2e\hbar v_F^2|n|B)^{1/2}$. Also, SdH[30] as well as IR transmission experiments[31] have revealed the existence of DFs in FLG samples.

## 4. Extrinsic Versus Intrinsic Differences Between Few-Layer-Thick and Bulk Graphite: Concluding Remarks

The weak coupling between graphene layers in bulk graphite is responsible for its quasi-2D behavior. It is reasonable to believe that there should be an intrinsic difference between graphene and graphite because of the weak- but not zero-coupling between two or three layers,[32] which provide extra, specific contributions to the band structure at certain k-positions in the BZ and a mixture of massive and massless QPs. However, experimental results obtained from conductivity and particularly magnetization measurements provide evidence for the existence of both types of QPs in bulk graphite, and hence it appears unlikely that only two or three specific layers contribute to the measured signal(s). We believe that it is too premature to speculate on the contributions of two-, three-, or *x*-layers of bulk graphite on any transport property because of the unclear role of lattice defects, as well as the uncertain influence of the weak coupling between graphene layers. Systematic experimental work is necessary to clarify this point. In the absence of these results, the separation of intrinsic and non-intrinsic features of bulk graphite as well as FLG will remain elusive.

One consequence of the Dirac spectrum given by the energy $E_n[K] \approx \pm 420|n|^{1/2}(B[T])^{1/2}$ and valid for graphite/graphene is that quantum effects should be observable at room temperature, since $\Delta E_n \approx 420$ K between n = 0 and n = ± 1 LL at an applied magnetic field $B$ = 1 T. Figure 2 shows the de Haas-van Alphen (dHvA) oscillations in the magnetization measured for a HOPG sample at different temperatures. Quantum oscillations at $B \sim 1$ T are clearly observable at 300 K, which is in agreement with theoretical predictions, suggesting that graphite may be a suitable material for quantum devices working under normal conditions. In comparison, the QHE is observed at room temperature in graphene only at a field of ~ 30 T,[15] and thus it seems clear that the quenched disorder in graphene essentially limits the practical use of this material.

If we define a metal as a material that effectively screens an applied electric field (*E*), then graphite should not be considered a good metal since *E* applied normal to the graphene layers can penetrate tens of nanometers, which is very different from typical metals where the field is screened within the first atomic layers. Theoretical predictions,[33-35] as well as recently performed experiments on graphite samples that are ca. 50 nm thick,[36] provide support for this important, but hitherto unnoticed behavior. This implies that the non-negligible influence of a bias voltage will enhance the possible applications of low-defect-density multilayer graphene samples.

The above discussion has focused primarily on the field dependence of the Hall signal, and we have clarified that in FLG as well as bulk high-quality graphite samples the Hall signals are basically consistent. One question that remains pertains to the field dependence of the longitudinal resistivity $\rho_{xx}(B)$. A comparison of the published data for graphite[3,4] and FLG/graphene samples[9-11] shows that in the latter samples, the magnetoresistance, defined as $\rho_{xx}(B \sim 1 \text{ T})/\rho_{xx}(0)$, is negligible in comparison with the ca. 1000% measured for bulk graphite at $T \sim 10$ K.[4] Recent work[37] suggests that this reproducible experimental observation is related to the lateral size of the sample or the distance between the voltage contacts, indicating the extraordinary large coherence or Fermi wavelength and the mean free path of the QPs in graphite.

Finally, it is worth exploring the possible use of multilayer graphene in spintronic devices. We speculate that multilayer graphene will have a large spin diffusion length, a property that can be used to transfer electron spins without losses between magnetic electrodes. In fact, recent studies on graphene samples with relatively low mobility indicate spin relaxation lengths in the micrometer range at room temperature.[38] Owing to the huge mobility obtained for high-quality graphite ($\mu \sim 10^6$ cm$^2$V$^{-1}$s$^{-1}$), we expect much longer spin relaxation lengths in multilayer graphene samples. On the other hand, graphite itself can be made ferromagnetic with a surprisingly high Curie temperature $T_c$ > 300 K, as revealed by research performed over the last few years.[39] There is no doubt that defects are responsible for this phenomenon, although the details are still under investigation. The possibility of obtaining ferromagnetic multilayer graphene with a Curie temperature above room temperature exhibiting large magnetoresistance is certainly an aim worth striving for.

We thank N. García and Igor Luk'yanchuk for enlightening discussions on the physics of graphite.

This work was supported by FAPESP and CNPq in Brazil, the DFG in Germany, and the EU under "Ferrocarbon".

[*] Y. Kopelevich
   Instituto de Física "Gleb Wataghin", Universidade Estadual de Campinas,
   Unicamp  13083-970, Campinas, São Paulo (Brazil); E-mail: kopel@ifi.unicamp.br

[**] P. Esquinazi
    Division of Superconductivity and Magnetism, Institute for Experimental
    Physics II University of Leipzig, Linnéstrasse 5, D-04103 Leipzig (Germany);
    E-mail:   esquin@physik.uni-leipzig.de

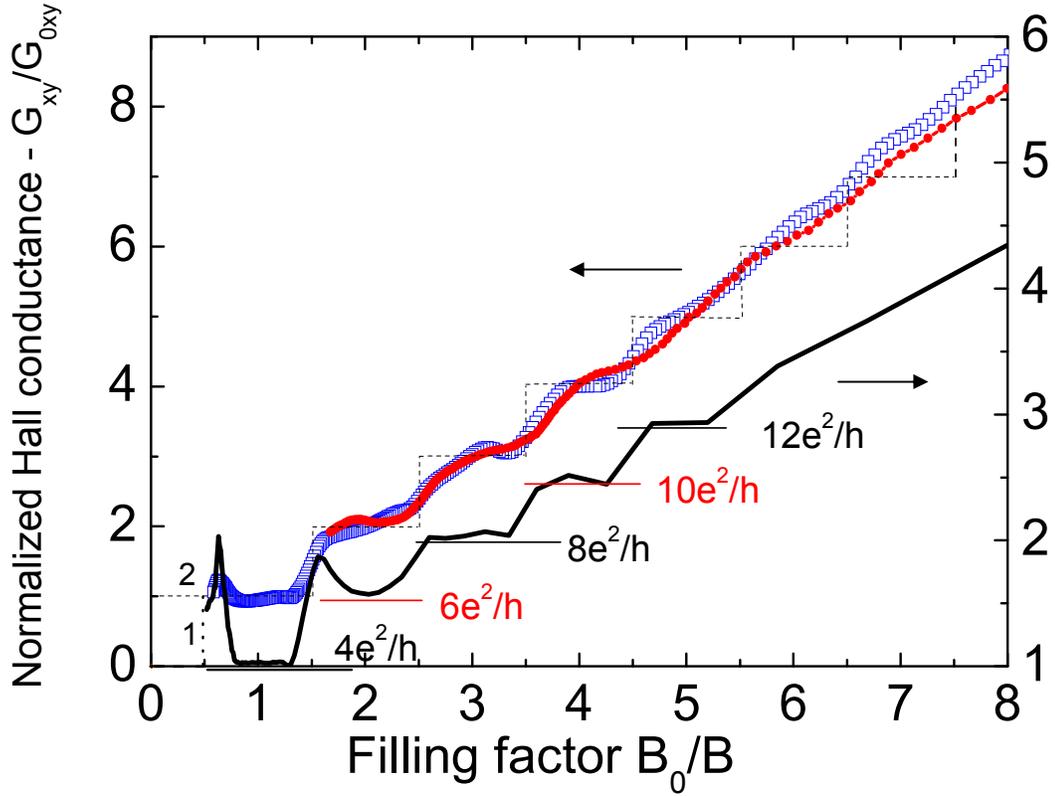

**Figure 1.** Normalized Hall conductance $G_{xy} = 1/R_{xy}$ obtained for two HOPG samples (HOPG-3 (□, left axis) [4] and HOPG-UC (continuous line, right axis)) plotted versus the inverse of the applied field. This quantity is written in terms of the filling factor $B_0/B$ for normalization purposes. The normalization field $B_0$ is obtained from the Shubnikov-de Haas oscillations in $R_{xx}$. For the HOPG samples, $B_0 = 4.68$ T. The staircases 1 and 2 correspond to the quantization steps expected for massive QPs with a conventional spectrum and for a two-parabola spectrum of bilayer graphite, respectively. The steps drawn at larger filling factors are only a guide and correspond to the expected plateaus positions according to theoretical predictions [21,25]. The data for HOPG-3 and HOPG-UC samples demonstrate the contributions from massive as well as massless carriers in both samples but with different weights. The expected absolute values of the Hall conductance for DFs (red markers) agree with the observed steps in sample HOPG-UC. The data points (●) correspond to the normalized data obtained for FLG from Novoselov et al. at a bias voltage of 80 V [9]. The normalization field $B_0 = 20$ T in this case, which corresponds to a larger electronic density as compared to the values for bulk HOPG samples. Similar to HOPG-3, massive QPs dominate the Hall signal in this two- or three-layer-thick sample.

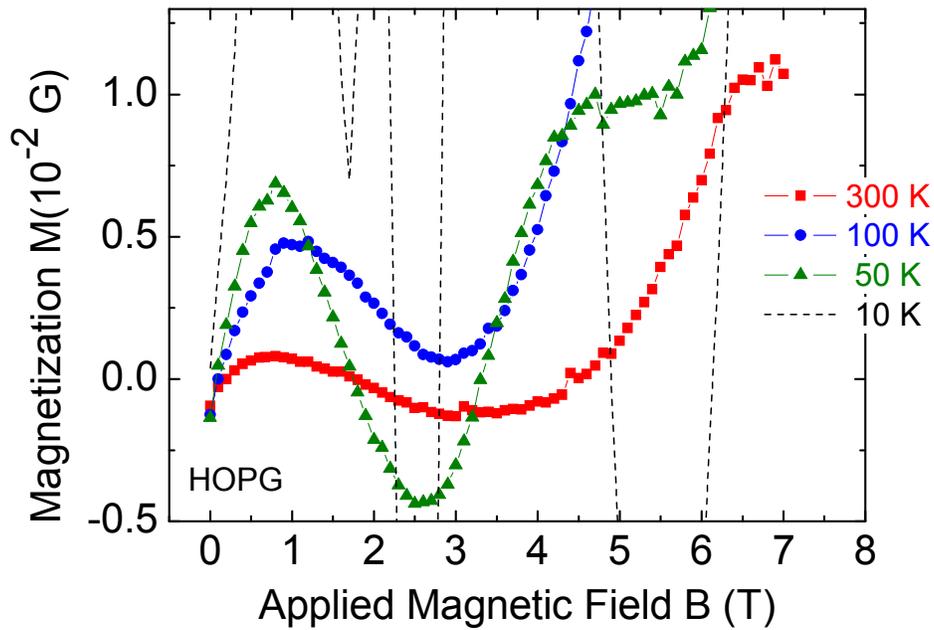

**Figure 2.** de Haas-van Alphen oscillations in the magnetization as a function of the applied field normal to the graphene planes for a HOPG sample measured at different temperatures. Note that these oscillations are observed up to room temperature, demonstrating the large LL separation expected for the Dirac-like spectrum. The magnetization values are obtained after subtracting the non-oscillating diamagnetic background signal from the measured curve.